# Magnon-magnon interaction and magnon relaxation time in ferromagnetic $Cr_2Ge_2Te_6$ monolayer

Ke Wang[1,2], Xiansong Xu[3], Yuan Cheng[2], Min Zhang[1], Jian-Sheng Wang[4], Hai Wang[1]*, and Gang Zhang[2]*

1) Xidian University, No. 2 Taibai Road, Xi'an, Shaanxi Province 710071, China

2) Institute of High Performance Computing, A*STAR, 138632, Singapore

3) Science, Mathematics and Technology Cluster, Singapore University of Technology and Design, 487372, Singapore

4) Department of Physics, National University of Singapore, 117551, Singapore

Email: wanghai@mail.xidian.edu.cn; zhangg@ihpc.a-star.edu.sg

## ABSTRACT

Despite the intense amount of attention and huge potential of two-dimensional (2D) magnets for applications in novel magnetic, magneto-optical, magneto-thermal and magneto-electronic devices, there has yet to be a robust strategy developed to systematically understand magnon-magnon (MMI) interactions at finite temperature. In this paper, we present a first-principles theoretical method to introduce the finite temperature magnon-magnon interaction into Heisenberg Hamiltonian through a nonlinear correction energy. The Wick's theorem is used to decouple the four-magnon operators to two-magnon order. We demonstrate the capabilities of this method by studying the strength of MMI in $Cr_2Ge_2Te_6$ (CGT) monolayer. The spin wave spectrum at finite temperature and the time-dependent spin autocorrelation function are explored. It is found that the magnon relaxation time due to magnon-magnon scattering increases with temperature because of the reduction in magnon energy, while decreases with wavevector and external magnetic field. Our results provide a new insight to understand the magnon damping and energy dissipation in two-dimensional ferromagnetic materials.

# 1. Introduction

The two-dimensional (2D) materials exfoliated from various bulk van der Waals sheets (vdW) attract substantial attention in materials science, condensed matter physics, and electronic engineering, because of their fascinating properties[1-4]. Although all major electronic classes (metals, insulators, and semiconductors) have been observed in 2D materials[5-7], 2D magnet was missing until the discovery of intrinsic ferromagnetism in Cr trihalide and chalcogenides in 2017[8,9]. Subsequently, numerous 2D magnets with long-range order have been discovered and investigated, such as $Fe_{5-x}GeTe_2$[10,11], $XPS_3$ (X=Cr, Mn, Fe)[12-16], $CrPS_4$, $Cr_2S_3$[17,18], $CoGa_2X_4$ (X= S, Se, Te)[19], $XH_2$ (X = Sc, Ti, V, Cr, Fe, Co, Ni) [20], CrOX (X=Cl, Br)[21,22], non-vdW transition-metal oxides[23] and some 2D transition metal carbides and nitrides (MXenes)[24,25].

For 2D magnets, magnetic anisotropy including the easy axis, the Kitaev interaction and the single ion anisotropy is of vital importance because sizeable MAE is beneficial to the long-range ferromagnetic order at finite temperature[26], so numerous researchers have been devoted to related study[27-31]. Based on the magnetic anisotropy, the Curie/ Neel temperature is predicted through the mean field theory. Meantime, several studies have been implemented to enhance the Curie/ Neel temperature for the application of 2D magnets in magneto-optical, magneto-thermal and magneto-electronic devices by functionalization[32-34], external field[8,35] and strain[36,37]. Below the Curie/Neel temperature, the time-resolved magneto-optical Kerr effect (TR-MOKE) and the time-resolved Faraday rotation (TRFR) are often employed to investigate the magnetic-optical coupling[9,38,39], demonstrating the rich magnetic behaviors in 2D magnets. The Hall effect (including spin Hall effect, anomalous Hall effect and thermal Hall effect) in 2D magnets also attracts substantial attention, which not only promotes the development of novel Hall devices but also provides a new route to detect the existence of magnetism[40-44]. Among these Hall effects, the thermal Hall effect is related to the coupling between spin wave and lattice vibration (magnon-phonon coupling)[45], and the magnon-phonon coupling is usually studied by Raman spectra[46,47] phonon spectrum[48,49] and spin-wave spectrum[50,51]. In spintronics, the manipulation of magnetic order also is a hot topic, and there are numerous manipulation strategies, such as stack[52,53], electrostatic doping[54,55], and pressure[56,57]. So far, despite intense study in this area, there is a lack of a robust theoretical method for studying the magnon-magnon interaction (MMI) at finite temperature, which is crucial for the coherent spintronics.

In this paper, we introduce the nonlinear MMI into Heisenberg Hamiltonian to describe the spin-wave spectrum of CGT monolayer at finite temperature. The influence of nonlinear MMI on magnon dispersion can be regarded as a nonlinear correction energy caused by the variety in magnon population. To observe the MMI, the time-dependence spin autocorrelation is employed, and the relaxation time $\tau_{MM}$ is obtained by fitting the spin autocorrelation with exponential function. We theoretically find the relaxation time $\tau_{MM}$ (contributed by MMI) increase with temperature but decreases remarkably with the external magnetic field. Our results shed a light on the understanding of dissipation of magnon energy at finite temperature.

## 2. Theoretical models and computational details

Here, the ferromagnetic system is described by the Heisenberg Hamiltonian including the nearest-neighbor (N) exchange and Zeeman energy:

$$H = -\frac{1}{2} \sum_{\substack{l,f \in N \\ l \neq f}} J\, \boldsymbol{S}_l \cdot \boldsymbol{S}_f - \sum_l g\mu_B \boldsymbol{B} \cdot \boldsymbol{S}_l. \tag{1}$$

For a given magnetic lattice, $\boldsymbol{S} = (S^x, S^y, S^z)$ is the spin vector whose amplitude is $S_0$. $J$ is the exchange constant of the nearest-neighbor interaction, $g$ is Landé factor, $\mu_B$ is the Bohr magneton, and $\boldsymbol{B}$ represents an external magnetic field along the $c$-axis. Without loss of generality, the equilibrium magnetization is assumed to parallel to the external field. In addition, transverse components $S^\pm = S^x \pm i S^y$ are defined to eliminate the dependence between $S^x$ and $S^y$. According to Holstein-Primakoff (HP) approximation[58,59], $S^{\pm}$ and $S^z$ for a given site can be written as

$$S^+ = \left(\sqrt{2S_0 - a^+ a}\right)a, \quad S^- = a^+\left(\sqrt{2S_0 - a^+ a}\right), \quad S^z = (S_0 - a^+ a), \tag{2}$$

where $a^+$ and $a$ are the creation and annihilation operators of magnon for the given site, respectively. The operators of magnon described in coordinate space are transferred into reciprocal space by the Fourier transform:

$$a_k = N^{-\frac{1}{2}}\sum_r e^{-i\bm{k}\cdot\bm{r}} a, \quad a_k^+ = N^{-\frac{1}{2}}\sum_r e^{i\bm{k}\cdot\bm{r}} a^+, \tag{3}$$

where $\bm{k}$ is the wave vector, $\bm{r}$ is the position vector of the lattice point, and $N$ is the number of unit cells in the supercell. The expectation value for local spin $\langle a_k^+ a_k \rangle$ can be calculated by the Planck distribution function:

$$\langle a_k^+ a_k \rangle = \frac{1}{\left(e^{\hbar\omega_k/k_B T} - 1\right)}, \tag{4}$$

with Boltzmann constant $k_B$ and the reduced Planck constant $\hbar$. At low temperature (close to zero), $\hbar\omega/k_B T$ tends to be infinite, resulting in an ignorable $\langle a_k^+ a_k \rangle$. With temperature increases, $\langle a_k^+ a_k \rangle$ increases significantly, thus cannot be ignored.

**2.1 Zero temperature model**.

For the completeness, firstly we introduce the spin spectrum model at zero temperature, at which the population of excited magnon is low, therefore the expectation value for local spin can be negligible with respect to $2S_0$[60,61]. The transverse components of spin vector for a given site can be rewritten as:

$$S^+ = \left(\sqrt{2S_0 - a^+ a}\right)a \approx \left(\sqrt{2S_0}\right)a, \quad S^- = \left(\sqrt{2S_0 - a^+ a}\right)a^+ \approx a^+\left(\sqrt{2S_0}\right). \tag{5}$$

The Hamiltonian can be given as:

$$H = H_0 + H_B = -\frac{1}{2} J \sum_{\substack{l,f \in N \\ l \neq f}} \left[ (S_0 - a_l^+ a_l)(S_0 - a_f^+ a_f) + S_0 (a_l a_f^+ + a_l^+ a_f) \right] - \sum_l g\mu_B B \cdot (S_0 - a_l^+ a_l), \quad (6)$$

Here $H_0$ and $H_B$ are:

$$H_0 = -\frac{1}{2} J \sum_{\substack{l,f \in N \\ l \neq f}} \left[ (S_0 - a_l^+ a_l)(S_0 - a_f^+ a_f) + S_0 (a_l a_f^+ + a_l^+ a_f) \right], \quad (7)$$

$$H_B = -\sum_l g\mu_B B \cdot (S_0 - a_l^+ a_l). \quad (8)$$

Following Eq. (3), Eq. (6) can be transferred into the reciprocal space. Then, the Hamiltonian in the reciprocal space can be written as:

$$H = H_0 + H_B = -\frac{1}{2} J \sum_{\substack{l,f \in N \\ l \neq f}} \left( S_0^2 - S_0 N^{-1} \sum_k (a_{l,k}^+ a_{l,k} + a_{f,k}^+ a_{f,k} - a_{f,k}^+ a_{l,k} \gamma_k - a_{l,k}^+ a_{f,k} \gamma_k) \right) - \sum_l g\mu_B B \cdot \left( S_0 - N^{-1} \sum_k a_{l,k}^+ a_{l,k} \right) \quad (9)$$

with $\gamma_k = \frac{1}{Z} \sum_{l,f \in N} e^{ik \cdot (r_f - r_l)}$. Z is the coordination number for the nearest interaction.

In each unit cell of monolayer CGT, it includes two magnetic $Cr^{3+}$ ions, then the Hamiltonian is:

$$H = H_0 + H_B = E_0 + \sum_k \hbar \omega_k^\pm a_k^+ a_k = E_0 + \sum_k \left( JZS_0 (1 \pm \gamma_k) + 2g\mu_B B \right) a_k^+ a_k, \quad (10)$$

where $E_0$ is the energy of ground state:

$$E_0 = -2NJZS_0 - 2Ng\mu_B B S_0. \quad (11)$$

In Eq. (10), the optical ($\hbar\omega^+$) and acoustic ($\hbar\omega^-$) branches of spin waves are gained when the plus and minus signs are taken, respectively.

**2.2 Magnon-magnon interaction at finite temperature.**

From Eq. (4), it is obvious that the expectation value for local spin increases with temperature because the excitation of substantial magnons, so that at finite temperature the transverse components of spin vector for a given site should be written as follows[62]:

$$S^+ = \left(\sqrt{2S_0 - a^+ a}\right) a \approx \sqrt{2S_0}\left(1 - \frac{a^+ a}{4S_0}\right) a, \quad S^- = a^+ \left(\sqrt{2S_0 - a^+ a}\right) \approx a^+ \sqrt{2S_0}\left(1 - \frac{a^+ a}{4S_0}\right). \quad (12)$$

Then, the higher order terms induced by the nonlinear MMI are included in Heisenberg Hamiltonian:

$$H(T) \approx H_0 + H_B - \frac{1}{2}J \sum_{\substack{l,f \in N \\ l \neq f}} \left[ a_l^+ a_l \, a_f^+ a_f + \frac{a_l^+ a_l \, a_l a_f^+}{4} + \frac{a_f^+ a_f \, a_l a_f^+}{4} + \frac{a_l^+ a_f a_l^+ a_l}{4} + \frac{a_l^+ a_f a_f^+ a_f}{4} \right]. \quad (13)$$

Next, the normal coordinate is introduced into Hamiltonian by Eq. (3), and the Fourier transform for the four-order terms are shown as following:

$$a_l^+ a_l a_f^+ a_f = N^{-2} \sum_{k-q} e^{-i(k-q)r_l} a_{k-q}^+ \sum_k e^{ikr_f} a_k \sum_{k'+q} e^{-i(k'+q)r_f} a_{k'+q}^+ \sum_{k'} e^{ik'r_l} a_{k'} = N^{-2} \sum_{k,k',q} \gamma_{k-k'-q} a_{k-q}^+ a_{k'+q}^+ a_{k'} a_k, \quad (14)$$

$$a_l^+ a_l a_l a_f^+ = N^{-2} \sum_{k-q} e^{-i(k-q)r_l} a_{k-q}^+ \sum_k e^{ikr_l} a_k \sum_{k'+q} e^{-i(k'+q)r_l} a_{k'+q}^+ \sum_{k'} e^{ik'r_l} a_{k'} = N^{-2} \sum_{k,k',q} \gamma_{k-q} a_{k-q}^+ a_{k'+q}^+ a_{k'} a_k, \quad (15)$$

$$a_f^+ a_f a_l a_f^+ = N^{-2} \sum_{k-q} e^{-i(k-q)r_f} a_{k-q}^+ \sum_k e^{ikr_f} a_k \sum_{k'+q} e^{-i(k'+q)r_f} a_{k'+q}^+ \sum_{k'} e^{ik'r_l} a_{k'} = N^{-2} \sum_{k,k',q} \gamma_k a_{k-q}^+ a_{k'+q}^+ a_{k'} a_k, \quad (16)$$

$$a_l^+ a_f a_l^+ a_l = N^{-2} \sum_{k-q} e^{-i(k-q)r_l} a_{k-q}^+ \sum_k e^{ikr_l} a_k \sum_{k'+q} e^{-i(k'+q)r_l} a_{k'+q}^+ \sum_{k'} e^{ik'r_f} a_{k'} = N^{-2} \sum_{k,k',q} \gamma_k a_{k-q}^+ a_{k'+q}^+ a_{k'} a_k, \quad (17)$$

$$a_l^+ a_f a_l^+ a_l = N^{-2} \sum_{k-q} e^{-i(k-q)r_l} a_{k-q}^+ \sum_k e^{ikr_l} a_k \sum_{k'+q} e^{-i(k'+q)r_l} a_{k'+q}^+ \sum_{k'} e^{ik'r_f} a_{k'} = N^{-2} \sum_{k,k',q} \gamma_k a_{k-q}^+ a_{k'+q}^+ a_{k'} a_k, \quad (18)$$

$$a_l^+ a_f a_f^+ a_f = N^{-2} \sum_{k-q} e^{-i(k-q)r_l} a_{k-q}^+ \sum_k e^{ikr_f} a_k \sum_{k'+q} e^{-i(k'+q)r_f} a_{k'+q}^+ \sum_{k'} e^{ik'r_f} a_{k'} = N^{-2} \sum_{k,k',q} \gamma_{k-q} a_{k-q}^+ a_{k'+q}^+ a_{k'} a_k. \quad (19)$$

Then, the Hamiltonian at finite temperature in reciprocal space can be represented by

$$H(T) = H_0 + H_B - JZN^{-1} \sum_{k,k',q} a_{k-q}^+ a_{k'+q}^+ a_{k'} a_k \left( \gamma_{k-k'-q} - \frac{\gamma_{k-q}}{2} - \frac{\gamma_{k'}}{2} \right). \quad (20)$$

There is no analytical solution for the four magnon operators, so it is necessary to decompose the products of four magnon operators. Wick's theorem, an algebraic strategy, states that the product of operators is equal to the sum of all possible pairs of operators, and allows one to handle the reduction problem quite easily. Here we adopt the Wick's theorem to decouple the four-magnon operators to two-magnon order[63]:

$$a^+_{k-q}a^+_{k'+q}a_{k'}a_k = \left(a^+_{k-q}a^+_{k'+q}\right)\left(a_{k'}a_k\right)+\left(a^+_{k-q}a_{k'}\right)\left(a^+_{k'+q}a_k\right)+\left(a^+_{k-q}a_k\right)\left(a^+_{k'+q}a_{k'}\right). \tag{21}$$

Meantime, the low order Taylor approximation is used to simplify the Hamiltonian:

$$\left(a^+_{k-q}a_{k'}\right)\left(a^+_{k'+q}a_k\right)=\left[\left\langle a^+_{k-q}a_{k'}\right\rangle+\left(a^+_{k-q}a_{k'}-\left\langle a^+_{k-q}a_{k'}\right\rangle\right)\right]\cdot\left[\left\langle a^+_{k'+q}a_k\right\rangle+\left(a^+_{k'+q}a_k-\left\langle a^+_{k'+q}a_k\right\rangle\right)\right]. \tag{22}$$

Finally, the Hamiltonian at finite temperature can be written as:

$$H(T)=H_0+H_B+E_1+JZN^{-1}\sum_{k,k',q}\left(\left\langle a^+_{k-q}a_k\right\rangle a^+_{k'+q}a_{k'}+\left\langle a^+_{k'+q}a_{k'}\right\rangle a^+_{k-q}a_k+\left\langle a^+_{k-q}a_{k'}\right\rangle a^+_{k'+q}a_k+\left\langle a^+_{k'+q}a_k\right\rangle a^+_{k-q}a_{k'}\right)\left(\frac{\gamma_{k-q}}{2}+\frac{\gamma_{k'}}{2}-2\gamma_{k-k'-q}\right) \tag{23}$$

with $\sum_{k'}\gamma_{k-k'}=\gamma_k\sum_{k'}\gamma_{k'}$. $E_1$ is the static energy difference from the ground state. Consequently, the magnon dispersion at finite temperature can be represented as:

$$\hbar\omega^{\pm}_k(T)=JZS_0\left(1\pm\gamma_k\right)\left(1-\frac{1}{NS_0}\sum_{k'}(1-\gamma_{k'})\left\langle a^+_k a_{k'}\right\rangle\right)+2g\mu_B B. \tag{24}$$

Therefore, the effect of nonlinear MMI induced by temperature on spin-wave spectrum can be regarded as a correction term. This modification of the spin-wave spectrum can be carried out by a self-consistent procedure.

**2.3 Computational details**

In this paper, we used the experimental lattice constants ($a = b = 6.83$ Å) [64], and the geometrical structure is relaxed by the Vienna *ab initio simulation* package (VASP) [65] using 500 eV cutoff energy and generalized gradient approximation (GGA) with $U$ of 1 eV[49, 66]. A vacuum space of 16 Å is set along the out-of-plane direction, because it was predicted that CGT monolayer has an out-of-plane magnetic easy axis[49]. The convergence for the energy difference of self-consistent iterations and the Hellmann–Feynman force were $10^{-8}$ eV and 0.001 eV/Å, respectively. A $k$-point mesh of 5×5×1 was used for structural relaxation, while a 9×9×1 mesh was used for self-consistent calculation to obtain the energy of CGT unit cell under different magnetic configurations.

## 3. Results and discussion

### 3. 1 Spin-wave spectrum at finite temperature.

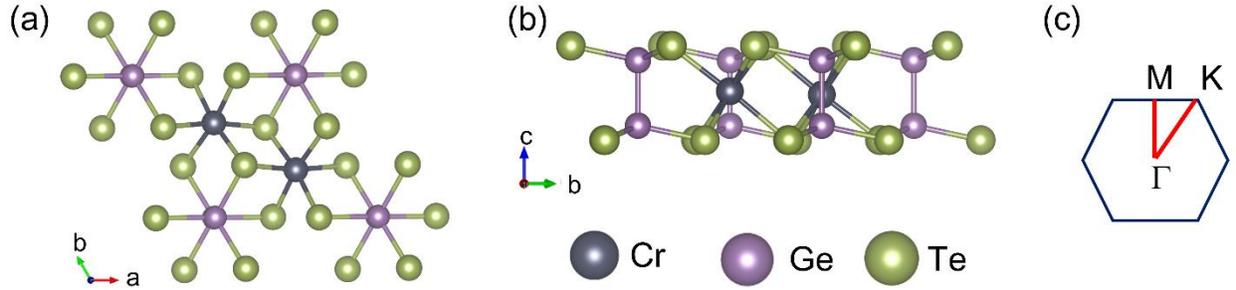

**Figure 1.** Top (a) and side (b) view of $Cr_2Ge_2Te_6$ (CGT) monolayer. The red parallelogram in (a) marks out the unit cell. The Cr, Ge and Te atoms are denoted in gray, purple and dark green colors, respectively. (c) The irreducible Brillouin zone of CGT monolayer.

The geometric structure and the path in Brillouin zone of CGT monolayer are shown in Figure 1. It was reported that Te atoms play a fundamental role in stabilizing the ferromagnetism of monolayer CGT through the super-exchange interactions [66-68], although they are nonmagnetic. In this work, only the nearest-neighboring interaction was taken into account, because the next-nearest-neighboring is one order of magnitude weaker[38,49]. The ferromagnetic ground state of CGT monolayer is confirmed, as the total energy of ferromagnetism (FM) phase is −48.407 eV, and that of antiferromagnetic (AFM) phase is −48.345 eV. The exchange constant can be calculated by:

$$J = (E_{AFM} - E_{FM})/(2ZS_0^2). \qquad (25)$$

In CGT monolayer, the coordinate number of nearest-neighboring interaction is 3. At zero temperature,

the exchange constant is 4.557 meV. In the following section, we would use the exchange constant $J$ at zero temperature to calculate the magnon dispersion and investigate the MMI, following the previous studies[69,70] because the temperature dependence of $J$ is neglectable. Moreover, because the Curie temperature $T_c$ of CGT monolayer is around 60 K [66,71,72], here we focus on the MMI at the finite temperature below 60 K. The magnon dispersion of CGT monolayer calculated through Eq. (16) at temperature of 0 K, 15 K, 30 K, and 45 K are represented by black, red, blue and green spheres in Figure 2(a). Due to the Zeeman energy in Heisenberg Hamiltonian, an external magnetic field of 0.1T along the $c$-axis is taken into consideration in spin-wave spectrum. It can be found that the magnon frequency at finite temperature is consistent globally with the zero temperature spectra except for a slight redshift. The temperature induced redshift in magnon dispersion at finite temperature should be ascribed to the nonlinear energy $\Delta E_{\text{nonlinear}}$ caused by the nonlinear MMI. And the occurrence of nonlinear MMI is resulted by the change in magnon population at finite temperature, so that the expectation value for local spin $\langle a_k^+ a_k \rangle$ cannot be ignored.

The correction energies $\Delta E_{\text{nonlinear}}$ induced by the nonlinear MMI for optical and acoustic magnons are presented in Figure 2(b) & 2(c). Here, we took the wave vectors ($k = 1.03 \times 10^6$, $1.03 \times 10^7$, $2.06 \times 10^7$, $3.09 \times 10^7$, $4.12 \times 10^7$ and $5.15 \times 10^7$ cm$^{-1}$) along the path from $\Gamma$ to M as examples. We can find that all the correction energies are negative revealing the influence of nonlinear MMI on spin-wave spectrum is to reduce the magnon frequency. The decrease in optical magnon is more pronounced than that in acoustic magnon, because the nonlinear correction energy is proportional to the magnon energy. More importantly, the nonlinear correction energies at all $k$ points become more negative with the increase of temperature, which agrees well with previous theoretical study on EuO using Dyson-Maleev theory[73].

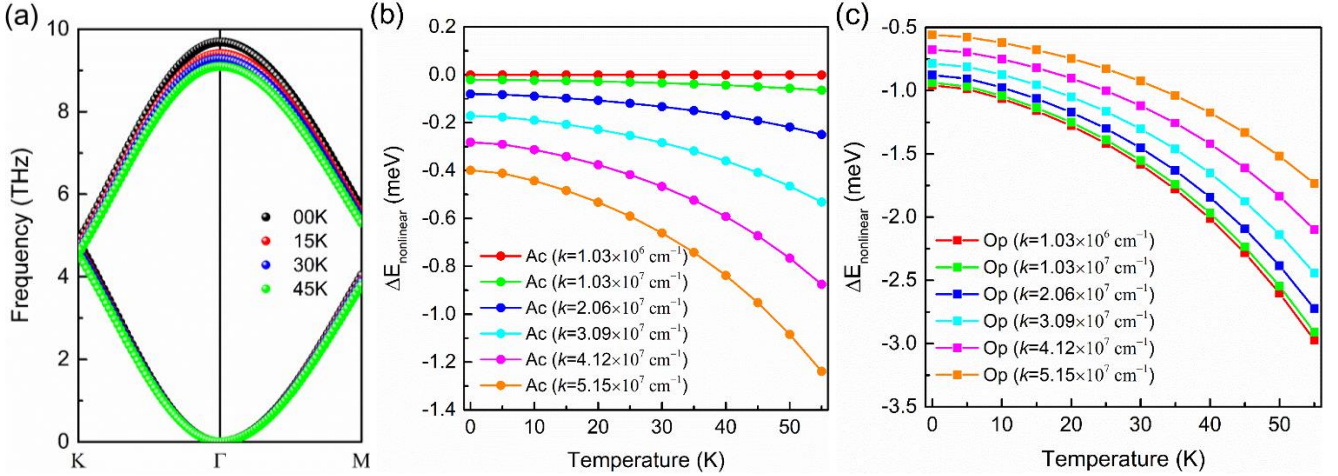

**Figure 2**. (a) Effects of temperature on spin wave spectrums. The nonlinear MMI correction energies for acoustic (b) and optical (c) magnons in CGT monolayer at finite temperatures. In (b & c), wave vector $k$=1.03×10$^6$, 1.03×10$^7$, 2.06×10$^7$, 3.09×10$^7$, 4.12×10$^7$ and 5.15×10$^7$ cm$^{-1}$ are sampled along the path from $\Gamma$ to M in Brillouin zone.

### 3.2 Magnon-magnon interaction and spin autocorrelation function

In this work, the MMI strength is investigated by calculating the spin autocorrelation functions $<S_k^+(t)S_{-k}^-(0)>$, $<S_k^-(t)S_{-k}^+(0)>$, and $<S_k^z(t)S_{-k}^z(0)>$. These spin autocorrelation functions are supposed to dominate electric spin injection, chemical potential-driven transport, which is related to the spin Seebeck effect.[70,74,75] The solutions of Heisenberg equation of motion $i\hbar\frac{\partial a_k(a_k^+)}{\partial t} = \hbar\omega_k a_k(a_k^+)$ based on $H(T)$ for $a_k(t)$ and $a_k^+(t)$ are:

$$a_k(t) = a_k(0)e^{-i\omega_k t}, \quad a_k^+(t) = a_k^+(0)e^{i\omega_k t}. \tag{26}$$

Due to the existence of external magnetic field along *c*-axis, we would put particular emphasis on the spin autocorrelation function $<S_k^z(t)S_{-k}^z(0)>$. In the reciprocal space, $S_k^z$ can be represented as:

$$S_k^z = N^{-\frac{1}{2}}\sum_l e^{-i k \cdot r_l} S_l^z = N^{\frac{1}{2}} S_0 \delta_{\vec{k},0} - N^{-\frac{1}{2}} \sum_{k'} a_{k'}^+ a_{k'+k}. \tag{27}$$

And then, $<S_k^z(t)S_{-k}^z(0)>$ can be written as:

$$\langle S_k^z(t)S_{-k}^z(0)\rangle = \frac{\text{Tr}\left\{e^{-\beta H}\left[\left(N^{\frac{1}{2}}S_0\delta_{k,0} - N^{-\frac{1}{2}}\sum_{k'}a_{k'}^+(t)a_{k'+k}(t)\right)\left(N^{\frac{1}{2}}S_0\delta_{-k,0} - N^{-\frac{1}{2}}\sum_{k'}a_{k'}^+(0)a_{k'-k}(0)\right)\right]\right\}}{\text{Tr}\left\{e^{-\beta H}\right\}}$$

$$= \langle NS_0^2\delta_{k,0}\rangle - S_0\delta_{k,0}\left\langle\left(\sum_{k'}a_{k'}^+(t)a_{k'+k}(t) + \sum_{k'}a_{k'}^+(0)a_{k'-k}(0)\right)\right\rangle + N^{-1}\sum_{k',k''}\langle a_{k'}^+(t)a_{k'+k}(t)a_{k''}^+(0)a_{k''-k}(0)\rangle \quad (28)$$

$$= NS_0^2\delta_{k,0} - 2S_0\delta_{k,0}\sum_{k'}\langle a_{k'}^+(0)a_{k'+k}(0)\rangle + N^{-1}\sum_{k'}e^{-i(\omega_{k'+k}-\omega_{k'})t}\langle a_{k'}^+(0)a_{k'+k}(0)a_{k'+k}^+(0)a_{k'}(0)\rangle$$

Here, we also employed Wick's theorem to decouple these four magnon operators[63], and it is:

$$\langle S_k^z(t)S_{-k}^z(0)\rangle = \begin{cases} NS_0^2\delta_{k,0} - 2S_0\delta_{k,0}\sum_{k'}\langle n_{k'}\rangle + N^{-1}\sum_{k'}\left(2\langle n_{k'}\rangle^2 + \langle n_{k'}\rangle\right) & (k=0) \\ N^{-1}\sum_{k'}e^{-i(\omega_{k'+k}-\omega_{k'})t}\langle n_{k'}\rangle\langle n_{k'+k}+1\rangle & (k\neq 0) \end{cases} \quad (29)$$

with $<n_k> = <a_k^+a_k>$. It is worth noting that $<n_k>$ is mainly contributed by acoustic magnons with respect to optical magnons, because of the low population of the latter. The spin autocorrelation function in long-wavelength limit ($k = 5.15\times10^4$ cm$^{-1}$=0.001 π/a) at 55 K and 5 K, and under external magnetic field $B$ of 0.1 T are shown in Figure 3(a) & 3(b), while that in short-wavelength limit ($k = 5.15\times10^7$ cm$^{-1}$=1 π/a, with $T = 55$ K and $B = 5$ T) is presented in Figure 3(c). As shown in Figure 3(a), it is obvious that the spin autocorrelation function decays with time and gradually converges. In order to obtain the relaxation time $\tau_{MM}$ governed by magnon-magnon interaction, an exponential decay function ($f(t)=A\cdot\exp(-t/\tau_{MM})$) was employed to fit the envelope of spin autocorrelation function, as plotted by the red solid line in Figure 3. The relaxation time $\tau_{MM}$ of spin autocorrelation function can reflect the strength of MMI, and the shorter $\tau_{MM}$ corresponds to a stronger MMI.

Comparing Figure 3(a) with Figure 3(b), there is a significant temperature dependence in the decay rate of spin autocorrelation function when temperature $T$ decreases from 55 K to 5 K. Comparing Figure 3(a) with Figure 3(c), it can be found that the increase of wave vector $k$ from long-wavelength limit to short-wavelength limit causes substantial reduction in the relaxation time $\tau_{MM}$. Besides, when the external magnetic field $B$ increases from 0.1 T to 5 T, the decay rate of spin autocorrelation function also increases. Figure 3(d) presents the spin autocorrelation function at $k=5.15\times10^4$ cm$^{-1}$ for $T=55$ K and $B=0.1$ T, where only the contribution of acoustic magnons is taken into consideration. Comparing Figure 3(a) to Figure 3(d), similar results suggest the spin autocorrelation function is mainly raised from acoustic magnon-magnon interaction. Next, we systematically explore the dependence of the spin autocorrelation function and relaxation time $\tau_{MM}$ on wave vector $k$, temperature $T$ and magnetic field $B$.

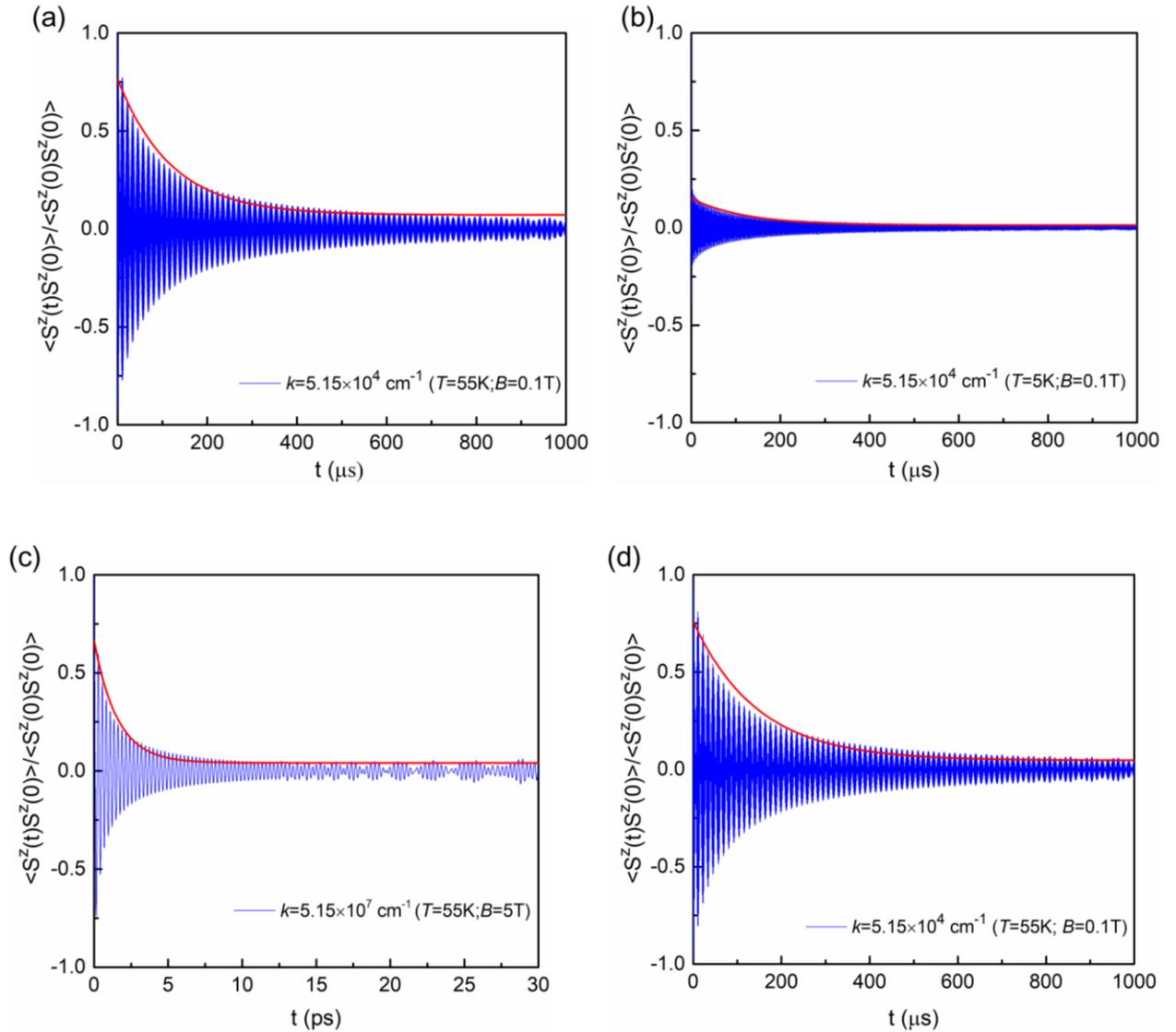

**Figure 3**. Autocorrelation functions at $k=5.15\times10^4$ cm$^{-1}$ for external magnetic field $B=0.1$ T and temperature $T=55$ K (a) or 5 K (b), and at $k=5.15\times10^7$ cm$^{-1}$ for $T=55$ K and $B=5$ T (c), while the spin autocorrelation function only contributed by acoustic magnons at $k=5.15\times10^4$ cm$^{-1}$ for $T=55$ K and $B=0.1$ T (d). The red solid lines represent the results fitted by exponential decay function.

### 3.3 Magnon relaxation time

In Figure 4, the wave vector dependence, the temperature dependence, and the frequency dependence of the spin autocorrelation function are presented. Firstly, we focus on the wavevector-dependence. Here, we extract the relaxation time $\tau_{MM}$ based on spin autocorrelation function throughout the whole wavelength regime, temperature is fixed at 30 K and magnetic field is 0.1 T, as shown in Figure 4(a). It is obvious that the relaxation time $\tau_{MM}$ decreases monotonously with wavevector $k$. The energy of acoustic magnon at the boundary of Brillouin zone is much higher than that at zone center, and the higher magnon energy leads to a faster decay rate of spin autocorrelation function. It is interesting that $\tau_{MM}$ at long-wavelength region and short-wavelength region show different dependences on wave vector $k$, which is bounded by $k=1.03\times10^6$ cm$^{-1}$.

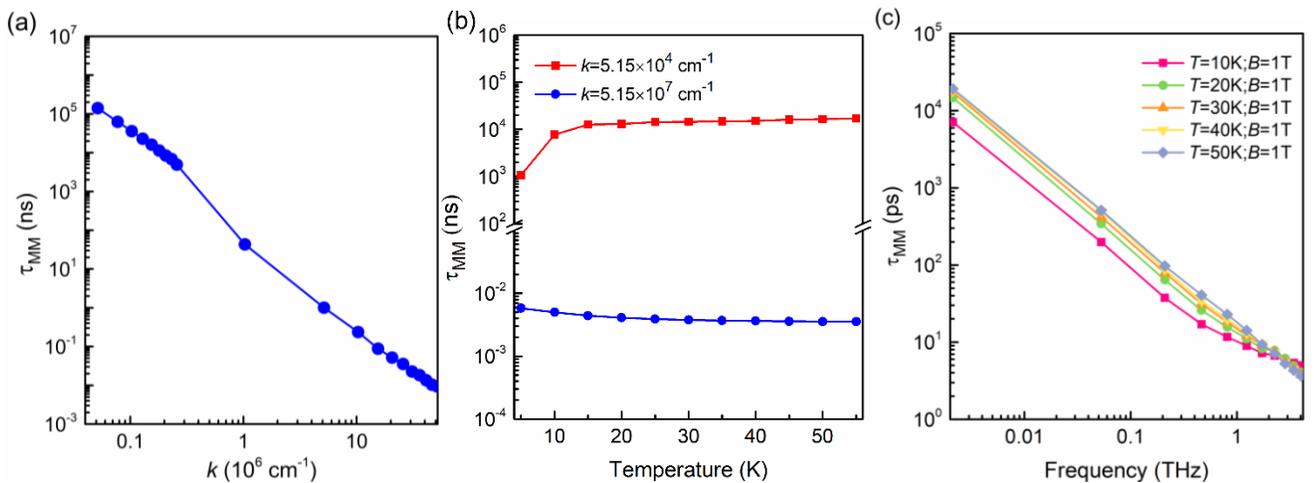

**Figure 4.** (a) The relaxation time $\tau_{MM}$ governed by magnon-magnon interaction at $T$=30 K and $B$=0.1T. The curve is plotted in log-log scale. (b) The temperature dependence of $\tau_{MM}$ at $k$=5.15×10$^4$ cm$^{-1}$ and $k$=5.15×10$^7$ cm$^{-1}$. Here $B$=1.0T. (c) $\tau_{MM}$ as a function of magnon frequency. Here $B$=1 T, and temperature changes from 10K to 50K.

The temperature dependence of relaxation time $\tau_{MM}$ for magnons in long-wavelength limit ($k$ = 5.15×10$^4$ cm$^{-1}$=0.001 $\pi$/a) and at the boundary of Brillouin zone ($k$ = 5.15×10$^7$ cm$^{-1}$=1 $\pi$/a) are presented in Figure 4(b). It is interesting that the relaxation times $\tau_{MM}$ in long-wavelength limit and short-wavelength limit show completely opposite temperature dependence. In long-wavelength limit, the relaxation time $\tau_{MM}$ increases with temperature, while it shows a reduction with temperature increasing at the boundary of Brillouin zone. It can be obtained from Eq. (29) that the decay rate of spin autocorrelation function depends on the magnon frequency. However, the magnon frequency decreases with temperature due to the nonlinear MMI, as shown in Figure 2. Therefore, the decay rate of spin autocorrelation function reduces with temperature, which leads to the increase in relaxation time $\tau_{MM}$ for long-wavelength magnons. At the boundary of Brillouin zone, relaxation time $\tau_{MM}$ decreases with temperature due to the strong anharmonic behavior. In Figure 4(c), we present the frequency dependence of relaxation time $\tau_{MM}$ under magnetic field $B$=1 T at different temperatures ($T$=10-50 K). It is obvious that the relaxation times $\tau_{MM}$ at different temperatures show similar frequency dependence. Meantime, an anharmonic behavior can be observed.

The dependences of relaxation time $\tau_{MM}$ on the strength of magnetic field at $k$ = 5.15×10$^4$ cm$^{-1}$ and $k$ = 5.15×10$^7$ cm$^{-1}$ at $T$=30 K are presented in Figure 5(a). It is found that the relaxation time in both long-wavelength and short-wavelength limit decreases with the enhancement of magnetic field, which is opposite with temperature dependence. Such an enhancement of the magnetic field increases the magnon frequency $\omega_k$, resulting in the rapid decay of spin autocorrelation function. This opposite impacts between temperature and magnetic field can also be observed in Figure 6, which shows the normalized magnetic moment per $Cr^{3+}$ ion at different temperature and magnetic field. The magnetic moment $M$ per $Cr^{3+}$ ion is calculated by[76]:

$$M(B,T) = g\mu_B \sum_l \left(S_0 - \langle a_l^+ a_l \rangle_{B,T}\right) = M_0\left(1 - \frac{1}{NS_0}\sum_k \langle a_k^+ a_k \rangle_{B,T}\right), \quad (30)$$

where $M_0 = g\mu_B S_0 = 3\mu_B$ is the magnetic moment of $Cr^{3+}$ at zero temperature. The dependence of normalized magnetic moment to $M_0$ on magnetic field $B$ at different temperatures is shown in Figure 6. It is obvious that the magnetic moment grows with the enhancement of magnetic field and tends to a specific value, while it decreases close to zero with the increase of temperature. These results reveal the expectation value of local spin $\langle a_k^+ a_k \rangle$ increases with temperature but decreases with the enhancement of magnetic field[77], which is in agreement with results in Figure 4(c) & Figure 5. Figure 5(b) presents $\tau_{MM}$ as a function of frequency under different magnetic field. Overall, $\tau_{MM}$ decreases with frequency increases, independent of the strength of magnetic field.

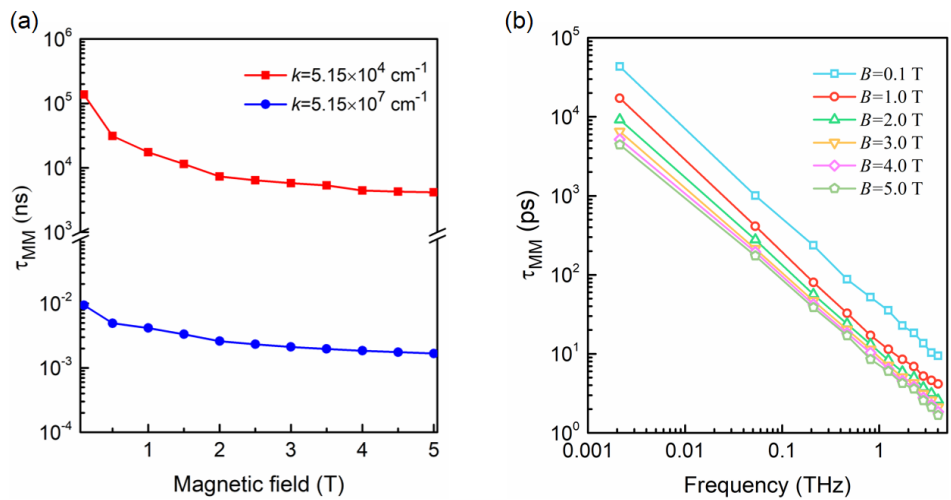

**Figure 5.** (a) The magnetic field dependence of $\tau_{MM}$ at $T$=30 K. (b) The frequency dependence of $\tau_{MM}$ at $T$=30 K at different magnetic field. In (b), the $\tau_{MM}$ versus frequency curves are shown in log-log scale.

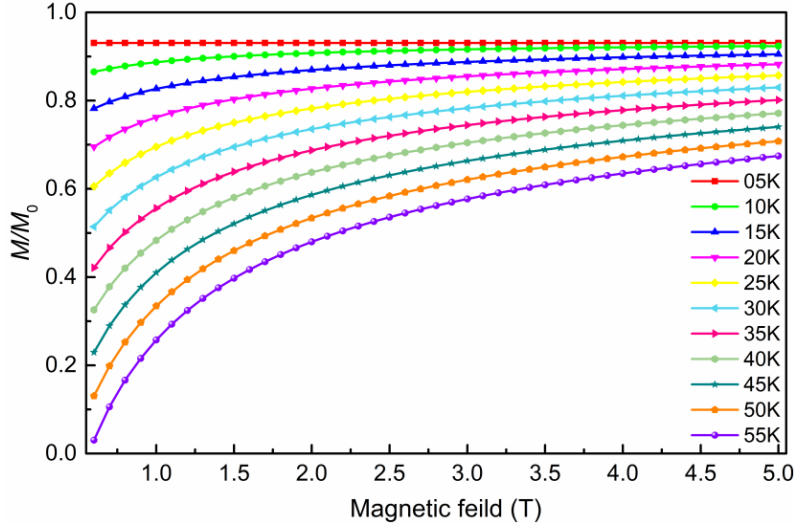

**Figure 6**. The magnetic moment on per $Cr^{3+}$ ion normalized to $M_0$ versus magnetic field. Temperature changes from 5K to 55K.

## 4. Conclusions

In summary, we established the Heisenberg Hamiltonian model for ferromagnetic 2D materials at finite temperature by taking the nonlinear magnon-magnon interaction into consideration. Based on this Hamiltonian model, it is found the increase of temperature results in a nonnegligible reduction in spin-wave spectrum, especially the optical branch. Furthermore, the relaxation time $\tau_{MM}$ governed by magnon-magnon interaction is also calculated based on decay of spin autocorrelation function. We find the relaxation time $\tau_{MM}$ increases with temperature but decreases with wave vector and magnetic field, because of the different magnon energy trends. All these results presented in our work are helpful for the utilization and manipulation of ferromagnetic 2D materials for applications in spintronic devices.

## ACKNOWLEDGEMENT

We gratefully acknowledge the financial supports from the Graduate School of Xidian University and the China Scholarship Council (grant no. 201906960032), and the use of computing resources at the Agency for Science, Technology and Research (A*STAR) Computational Resource Centre, Singapore.